\documentclass[11pt]{article}
\pdfoutput =1
\usepackage{amsmath}
\usepackage{verbatim}
\usepackage{amssymb}
\usepackage{mathrsfs}
\usepackage{inputenc}
\usepackage{textcomp}
\usepackage{appendix}
\usepackage{authblk}
\usepackage{amsfonts}
\usepackage{amssymb}
\usepackage{makeidx}
\usepackage{graphicx}
\usepackage{slashed}
\usepackage{wrapfig}
\usepackage{subcaption}
\usepackage{tikz}
\newcommand{\e}{\epsilon}
\newcommand{\be}[1]{\begin{equation}\label{#1} }
\newcommand{\ee}{\end{equation}}
\newcommand{\bea}[1]{\begin{eqnarray}\label{#1} }
\newcommand{\eea}{\end{eqnarray}}
\newcommand{\p}{\partial}
\newcommand{\refb}[1]{(\ref{#1})}

\renewcommand{\L}{{\mathcal{L}}}

\newcommand{\Q}{\mathcal{Q}}

\newcommand{\bL}{\bar{{\mathcal{L}}}}
\newcommand{\bQ}{\bar{\mathcal{Q}}}

\renewcommand{\>}{\rangle}

\newcommand{\w}{\omega}
\newcommand{\tw}{\tilde{\omega}}

\newcommand{\eps}{\varepsilon}

\renewcommand{\a}{\alpha}
\newcommand{\ta}{\tilde{\alpha}}
\newcommand{\tb}{\tilde{b}}

\newcommand{\C}{\tilde{C}}

\renewcommand{\t}{\tau}
\newcommand{\s}{\sigma}

\begin{document}
\date{}

\title{\textbf{Exotic Origins of Tensionless Superstrings}}

\author[1]{Arjun Bagchi}
\author[2]{Aritra Banerjee}
\author[3]{Shankhadeep Chakrabortty}
\author[1]{Pulastya Parekh}
\affil[1]{\small{Indian Institute of Technology Kanpur, Kalyanpur, Kanpur 208016. INDIA.} \\}
\affil[2]{CAS Key Laboratory of Theoretical Physics, Institute of Theoretical Physics, Chinese Academy of Sciences,  Beijing 100190, CHINA.\\}
\affil[3]{Indian Institute of Technology Ropar, Rupnagar, Punjab 140001, INDIA.\\}
\affil[ ]{Email: abagchi@iitk.ac.in, aritra@itp.ac.cn, s.chakrabortty@iitrpr.ac.in, pulastya@iitk.ac.in}

\maketitle

\begin{abstract}
A new type of tensionless superstring theory, called the Inhomogeneous tensionless superstring, has been recently introduced. This is characterised by a residual symmetry algebra on the worldsheet richer in structure than the previously known symmetry algebra related to Homogeneous tensionless superstring. In this paper, we investigate theories of tensile superstring that the Inhomogeneous tensionless superstring with manifestly real fermions could arise from. We provide two different candidates for these parent tensile theories. The nature of the limit dictates that these theories have some exotic features.  
\end{abstract}

\section{Introduction}
String theory presently is the most viable of the formulations for a theory of quantum gravity. The theory is endowed with a characteristic length-scale $\alpha'$, -- the length of the fundamental string. In the limit $\alpha' \to 0$, the string reduces to a point particle and we recover known physics, i.e. low-energy supergravity, making it possible to explore the classical regime of gravity from string theory. The opposite limit, $\alpha' \to \infty$ is expected to explore the extreme quantum sector of gravity. 

The tension of the string is proportional to the inverse of $\alpha'$ and hence the $\alpha' \to \infty$ is the limit where the tension of the string goes to zero. This tensionless limit has been of interest since \cite{Schild:1976vq} due to a variety of reasons. In this limit, the masses of all the string states go to zero and it is expected that there would be an emergent symmetry 
much larger than the usual symmetries of string theory. This has been explored in the early days by Gross and Mende from the point of view of string scattering \cite{Gross:1987kza,Gross:1987ar, Gross:1988ue}, and of late has been related to higher spin holographic dualities \cite{Gaberdiel:2010pz, Chang:2012kt, Gaberdiel:2014cha}. Another, less explored but possibly more intriguing connection, is with the physics of Hagedorn transitions \cite{Atick:1988si}. It was noted in \cite{Pisarski:1982cn, Olesen:1985ej} that the tension of the string decreases to zero as it approaches the Hagedorn temperature. 

Our previous construction of tensionless strings has focussed on utilising the power of symmetries on the worldsheet. In \cite{Bagchi:2013bga}, it was shown that the 3d Bondi-Metzner-Sachs (BMS$_3$) algebra or equivalently the 2d Galilean Conformal Algebra, arises on the worldsheet of the tensionless string as residual gauge symmetries in the equivalent of the conformal gauge. This is principally because the tensionless string is the 1-d extended analogue of the massless point particle and the metric on the worldsheet becomes null. A limiting procedure on the worldsheet was proposed which took one from the symmetries of the tensile worldsheet (Virasoro $\otimes$ Virasoro) to the BMS algebra. In \cite{Bagchi:2015nca}, aspects of the bosonic tensionless theory were further developed and the power of the mapping on the worldsheet was further clarified. The move to superstrings threw open new and interesting possibilities, which we briefly review below. The most striking of this was the discovery of a completely new tensionless superstring, which we will refer to as the Inhomogeneous tensionless superstring \cite{Bagchi:2017cte}, as opposed to the Homogeneous tensionless superstring described earlier in  \cite{Lindstrom:1990ar, Gamboa:1989px, Gamboa:1989zc} and also understood via limiting procedures in  \cite{Bagchi:2016yyf}.

An intriguing connection between tensionless superstrings and ambitwistors has recently been unearthed \cite{Casali:2016atr}. The version of the tensionless superstring that has been used is the Homogeneous one. But our formulation of the Inhomogeneous version points to the tantalising prospect of the discovery of a richer variant of the ambitwistor string. A very important question in the context of ambitwistors is tracing their origin to a parent superstring. Some attempts have been made in \cite{Lee:2017utr}. Keeping this in mind, and also with an eye out to applications of tensionless strings to the physics of the Hagedorn phase transitions, in the present paper, we attempt to answer the question what parent theory our tensionless strings arise from. We will be mostly focused on our newly discovered Inhomogeneous version in the current note.

\section{Tensionless Superstrings}

The tensionless equivalent of the RNS action \cite{Lindstrom:1990ar} for the spacetime bosonic fields $X^\mu$ and its supersymmetric partners $\psi^\mu$ is given by 
\be{1}
S=\int d^2\s \Big[V^a V^b \p_a X \cdot \p_b X + i\bar{\psi}\cdot\rho^a \p_a \psi \Big].
\ee
In the tensionless limit, the worldsheet metric becomes degenerate and is replaced by a product of two vector densities $V^a$:  
\be{}
T \sqrt{-g} g^{ab} \to V^a V^b, \quad \mbox{as} \quad T \to 0. 
\ee
The equivalent of the conformal gauge in tensile strings in this case is the choice $V^a=(1,0)$. In the action \refb{1}, $\rho^a$ are the 2d gamma matrices which now satisfy a modified Clifford algebra:
\be{gamma}
\{\rho^a,\rho^b\}=2V^aV^b.
\ee
The above equation \refb{gamma} is central to the formulation of different versions of tensionless superstrings. The equation admits two (and only two) inequivalent classes of solutions for $\rho^a$ that gives rise to the two inequivalent theories. 

\paragraph{Homogeneous Tensionless Superstrings:} The simplest possible choice is $\rho^a=V^a\mathbb{I}_{2\times 2}$, which in our gauge reads $\rho^0 = \mathbb{I}, ~\rho^1 = \mathbb{O}$. The action (\ref{1}) in this case becomes
\be{3}
S=\int d^2\s \Big[\dot{X}^2+ i\bar{\psi}\cdot\dot{\psi} \Big].
\ee
This describes what we call the \textit{Homogeneous Tensionless Superstring} \cite{Bagchi:2016yyf}. One can also consider the action in the worldsheet superspace \cite{Gamboa:1989zc},
where the generators of the symmetry transformations give rise to the following residual symmetry algebra: 
\bea{SGCAH}
&& [L_m,L_n]=(m-n)L_{m+n}, \qquad [L_m, M_n]=(m-n)M_{m+n}, \nonumber\\
&& [L_m,Q^\a_r]=\left(\frac{m}{2}-r\right)Q^\a_{m+r}, \qquad\{Q^\a_r,Q^{\a'}_s\}=\delta^{\a\a'}M_{r+s}.
\eea
Here the spinor index $\a$ denotes $\pm$. The bosonic indices $m,n$ take integer values while fermionic indices $r,s$ have half integer values. This algebra is the classical part of a supersymmetric extension of the (3d) Bondi-Metzner-Sachs (BMS) algebra. This algebra, where the two fermionic generators behave in a similar manner, is known as the homogeneous version of Super BMS (SBMS$_H$). This algebra has interestingly also arisen in the context of asymptotic symmetries of supergravity in 3d flat spacetimes \cite{Lodato:2016alv}. 

\paragraph{Inhomogeneous Tensionless Superstrings:}The other inequivalent class of the modified gamma matrices that solve the deformed Clifford algebra \refb{gamma} is given by \cite{Bagchi:2017cte}
\be{}
\rho^0=\begin{pmatrix} 1& 0 \\ 0 & -1\end{pmatrix} , \quad \rho^1=\begin{pmatrix} 0& 0 \\ 1 & 0\end{pmatrix}.
\ee
For this class of matrices, the associated fermions ($\psi^\mu$) need not be Majorana spinors. In \cite{Bagchi:2017cte} they were considered to be complex, however in this work we will work with real fermions. Considering the above choice of gamma matrices the action \refb{1} takes the form
\be{action_tless}
S=\int d^2\sigma \Big[\dot{X}^2+i(\psi_1\cdot\dot{\psi}_0+\psi_0\cdot\dot{\psi}_1-\psi_0\cdot\psi'_0)\Big],
\ee
where $\psi^\mu_0$ and $\psi^\mu_1$ are the components of the spinor $\psi^\mu$ which we can take to be real (Majorana).  The residual symmetry algebra associated to this action can also be calculated in a manner similar to \cite{Bagchi:2017cte}.  The symmetries that leaves the above action invariant comprises of local diffeomorphisms (paramterized by $\xi$) and supersymmtery transformations ($\e$) given by 
\begin{subequations}\label{Tsusy}
\bea{}
\delta_\xi\psi_0&=&\xi^a\p_a\psi_0+\frac{1}{4}\p_a\xi^a\psi_0, \quad  \delta_\xi\psi_1 = \xi^a\p_a\psi_1+\frac{1}{4}\p_a\xi^a\psi_1+\frac{1}{2}\p_1\xi^0\psi_0 \\
\delta_\e\psi_0&=&-\e^1\dot{X},\quad \delta_\e\psi_1=-\e^0\dot{X}-\e^0 X',~~
\delta_\xi X=\xi^a\p_aX,\quad\delta_\e X=i(\e^0\psi_0+\e^1\psi_1)~~
\eea 
\end{subequations}
where the parameters obey the conditions (Lorentz indices have been suppressed):
\bea{bhgb}
 \p_0\xi^0=\p_1\xi^1,\quad\p_0\xi^1=0, ~~
 \p_0\e^0=\p_1\e^1,\quad\p_0\e^1=0. 
\eea
These transformations can be realised on a modified superspace associated with the string worldsheet, parameterised by Grassmanian coordinates $\a$ and $\chi$. A general superfield $Y$ on this superspace can then be expanded as{\footnote{We should mention here, in general we could also have a term containing an auxiliary field of the form $\a\chi B(\s^a)$, however without any loss of generality we can put $B=0$ for our case.}}  
\bea{superspace_t}
Y(\sigma^a,\a,\chi)
&=&X(\sigma^a)+i\a\psi_0(\sigma^a)+i\chi\psi_1(\sigma^a).
\eea
Symmetry transformations on superspace coordinates are {\footnote{The Grassmanian parameters $\e^a$ as well as the coordinates $\a$ and $\chi$ are manifestly real here and above.}}:
\be{}
\delta\t=\xi^0+i(\e^0\chi+\e^1\a),\, \delta\s=\xi^1+i\e^1\chi, \, \delta\a=\e^0+\frac{1}{4}\p_a\xi^a\a+\frac{1}{2}\p_1\xi^0\chi, \,\delta\chi=\e^1+\frac{1}{4}\p_a\xi^a\chi. 
\ee
The differential generators of the above transformations read
\bea{}\label{inhomgen}
L_n&=&ie^{in\sigma}\Big[\p_\sigma+in\Big(\frac{\a\p_\a+\chi\p_\chi}{2}+\tau\p_\tau\Big)-\frac{n^2}{2}\tau\chi\p_\a\Big],\ M_n=ie^{in\sigma}\Big[\p_\tau+\frac{in}{2}\chi\p_\a\Big],  \nonumber \\
G_r&=&e^{ir\sigma}\Big[\p_\chi+i(\a\p_\tau+\chi\p_\sigma)+ir\tau\p_\a-r\chi\tau\p_\tau\Big],\quad\ \ \ H_r=e^{ir\sigma}\Big[\p_\a+i\chi\p_\tau\Big].
\eea
All of these generators are manifestly hermitian and close to form: 
\bea{agcai} 
&&[L_m,L_n]=(m-n)L_{m+n}, \quad [L_m,M_n]=(m-n)M_{m+n}, \nonumber \\
&&[L_m,G_r]=\Big(\frac{m}{2}-r\Big)G_{m+r}, \quad [M_m,G_r]=\Big(\frac{m}{2}-r\Big)H_{m+r},\ \{G_r,G_s\}=2L_{r+s},\nonumber \\
&&[L_m,H_r]=\Big(\frac{m}{2}-r\Big)H_{m+r}, \quad \{G_r,H_s\}=2M_{r+s}.
\eea
 This algebra clearly has more non-zero commutators and hence a richer structure than SBMS$_H$ \refb{SGCAH}. The fermionic generators ($G_r$, $H_r$) here behave differently to the Homogeneous case. Hence this avatar of the Super-BMS algebra is also called the Inhomogeneous SBMS or SBMS$_I$ and the associated string was called \textit{Inhomogeneous Tensionless Superstring}. This algebra has also been found to be the asymptotic symmetries of a twisted version of supergravity in 3d asymptotically flat spacetimes  \cite{Lodato:2016alv}. 
 
\medskip

\noindent The equations of motion obtained by varying \refb{action_tless} with respect to $X^\mu$ and $\psi^\mu$ reads:
\be{eomn}
\ddot{X}^\mu=0, \quad \dot{\psi}_0^\mu=0, \quad \dot{\psi}_1^\mu={\psi'}_0^\mu. 
\ee
We can solve these equations with closed string boundary conditions to obtain
\bea{NSsoln}
X^\mu(\tau,\sigma)&=&x^\mu+\sqrt{2c'}B^\mu_0\tau+i\sqrt{2c'}\sum_{n\neq 0}\frac{1}{n}(A^\mu_n-in\tau B^\mu_n)e^{-in\sigma},\nonumber \\
\psi_0^\mu(\sigma,\tau)&=&\sqrt{c'}\sum_r \beta^{\mu}_r e^{-ir\sigma}, ~~~
\psi_1^\mu(\sigma,\tau)=\sqrt{c'}\sum_r [\gamma^\mu_r-ir\tau\beta^{\mu}_r] e^{-ir\sigma}. 
\eea
Here, $n \in \mathbb{Z}$, $r \in \mathbb{Z} + \frac{1}{2}$ (we focus of the NS-NS sector). $c'$ is a constant with dimensions of (mass)$^{-1}$. The hermiticity of the generators must follow from the above mode expansions. This in turn implies the hermiticity of the modes $\gamma$ and $\beta$, i.e. 
$ \gamma^\dagger_{r}=\gamma_{-r},\quad\beta^\dagger_{r}=\beta_{-r}.$ It is straightforward to write down the generators in terms of oscillator modes, and check that the constraint algebra closes to SBMS$_I$ \refb{agcai}. 

Upon quantization, the commutators of the modes are given by
\be{abb}
[A_m^\mu,B_n^\nu] = 2m\delta_{m+n}\eta^{\mu\nu}, \quad \{\gamma_{r}^{\mu},\beta_{s}^{\nu} \} = 2\delta_{r+s}\eta^{\mu\nu}.
\ee
We would like to have an oscillator construction of modes hence transform to a new basis:
\be{sbasis}
C^\mu_n=\frac{1}{2}\Big(A^\mu_n+B^\mu_n\Big),\, \C^\mu_n=\frac{1}{2}\Big(-A^\mu_{-n}+B^\mu_{-n}\Big); \quad
\w^\mu_r=\frac{1}{2}\Big(\gamma^\mu_r+\beta^\mu_r\Big),\,\ \tilde{\w}^\mu_r=\frac{i}{2}\Big(-\gamma^\mu_{-r}+\beta^\mu_{-r}\Big).
\ee
These in turn follow the canonical commutation relations for harmonic oscillators:
\bea{ccr}
[C_m^\mu,C_n^\nu] &=&[\C_m^\mu,\C_n^\nu]= m\delta_{m+n}\eta^{\mu\nu},~~~
 \{\w_{r}^{\mu},\w_{s}^{\nu} \}= \{\tw_{r}^{\mu},\tw_{s}^{\nu} \}=\delta_{r+s}\eta^{\mu\nu}.
\eea
Since we demand $A, B, \gamma, \beta$ all to be manifestly hermitian, we find $\w_r$ is hermitian, $\tilde{\w}_r$ is anti-hermitiant: 
\be{wanti} \w^\dagger_r=\w_{-r};\quad \tilde{\w}^\dagger_r=-\tilde{\w}_{-r} \ee
While the commutation relations \refb{ccr} still hold, we also have the extra input,
$ \{\tw_r,\tw^\dagger_r\}=-1. $
This unconventional commutator allows us to define creation and annihilation operators in a consistent way. We will be able to define a vacuum $|00\>_C$ with the conditions 
\be{cvacc} C_n|00\>_C=\C_n|00\>_C=\w_r|00\>_C=\tw_r|00\>_C=0\ \ \  \forall\ n,r>0. \ee
Negative modes of the oscillators can be used to build up states on the vacuum in the usual way.

\section{Exotic Parent I: Maverick Superstrings}

Now that we have a well-defined tensionless theory, we will attempt to trace its origins to parent superstring theories. We will construct two different avenues of obtaining such theories. We first start with what we will call a maverick superstring theory. 

Let us consider the following gauge-fixed action for a tensile superstring with worldsheet bosonic fields $X$ and its supersymmetric partners $\psi$:
\be{taction}
S=\frac{1}{4\pi\a'}\int d^2\sigma \Big[2\p_+X\cdot\p_-X+i(\psi_{+}\cdot\partial_{-}\psi_{+}+\psi_{-}\cdot\partial_{+}\psi_{-})\Big].
\ee
Here $\psi_\pm$ are the two spinorial components of $\psi$.
We assume $\psi_+$ to be hermitian and $\psi_-$ to be anti-hermitian. We stress that this does not spoil the hermiticity of the action. Under local diffeomorphism ($\xi$) and supersymmetry transformations ($\e$)
\be{susy}
\delta_\xi X=\xi^a\p_aX, \,  \delta_\xi\psi_\pm = \xi^a\p_a\psi_\pm+\frac{1}{2}\p_\pm\xi^\pm\psi_\pm, \quad
\delta_\e X=i(\e^+\psi_++\e^-\psi_-), \, \delta_\e\psi_\pm = -2\e^\pm\p_\pm X. 
\ee 
leave the action invariant. The infinitesimal supersymmetry parameters $\e^\pm$ form a two-component spinor where $\e^-$ is anti-hermitian keeping $X$ and $\delta X$ real. Invariance of the action \refb{taction} also implies the consistency conditions on $\xi$ and $\e$:
$ \p_\pm\xi^\mp=\p_\pm\e^\mp=0$, which are same as in the usual tensile superstring theory. 
The superspace coordinates $\theta,\bar{\theta}$ 
also form a two component spinor where one component is anti-hermitian. The transformations in superspace that keep the action \refb{taction} invariant are the same as in the usual tensile string
and we can find the generators in a differential form:
\bea{tensgen}
\mathcal{L}_n&=&ie^{-in\sigma^+}\Big(\p_++\frac{in}{2}\theta\p_\theta\Big),\ \ \ \ \mathcal{Q}_r=e^{-ir\sigma^+}(\p_\theta+i\theta\p_+), \nonumber \\ 
\bar{\mathcal{L}}_n&=&ie^{-in\sigma^-}\Big(\p_-+\frac{in}{2}\bar{\theta}\p_{\bar{\theta}}\Big), \ \ \ \ \bar{\mathcal{Q}}_r=e^{-ir\sigma^-}(\p_{\bar{\theta}}+i\bar{\theta}\p_-). 
\eea
These close to form the usual Super-Virasoro algebra: 
\be{algv}
[\L_n, \L_m] = (n-m) \L_{n+m}, \quad [\L_n, \Q_r] = \left(\frac{n}{2} - r \right) \Q_{n+r}, \quad \{ \Q_r, \Q_s \} = 2 \L_{r+s}, 
\ee
(and similarly for the anti-holomorphic part). It is interesting to note that the anti-hermicity of the maverick superstrings does not manifest itself in the symmetry algebra, even though the generator $\bar{\mathcal{Q}}$ is anti-hermitian. 

The same algebra can also be obtained from the constraints. The solutions of the equations of motion $\p_-\p_+ X =0$ and $\p_{\pm}\psi_{\mp}=0$, can be constructed in terms of mode expansions in the NS-NS sector:
\begin{subequations}
\bea{}
X^\mu(\tau,\sigma)&=&x^\mu+2\sqrt{2\a'}\a^\mu_0\tau+i\sqrt{2\a'}\sum_{n\neq 0}\frac{1}{n}\Big[\a^\mu_ne^{-in(\tau+\sigma)}+\ta^\mu_ne^{-in(\tau-\sigma)}\Big], \\
\psi_+^\mu(\sigma,\tau)&=&\sqrt{2\a'}\sum_{r \in \mathbb{Z} + \frac{1}{2}} b^{\mu}_r e^{-ir(\tau+\sigma)}, \quad \psi_-^\mu(\sigma,\tau)=\sqrt{2\a'}\sum_{r \in \mathbb{Z} + \frac{1}{2}} \tilde{b}^{\mu}_{r} e^{-ir(\tau-\sigma)}.
\eea
\end{subequations}
The canonical commutation relation of $X^{\mu}$ and $\psi_{\pm}^{\mu}$ is:
\be{modecom1}
[X^\mu(\sigma),\dot{X}^\nu(\sigma')]=\eta^{\mu\nu}\delta(\sigma-\sigma'), \quad \{ \psi^\mu_\a(\sigma),\psi^\nu_{\a'}(\sigma') \}=\eta^{\mu\nu}\delta_{\a\a'}\delta(\sigma-\sigma').
\ee
The above brackets translate into (anti-)commutation relations between the modes, 
\be{mode}
[\a_m^\mu,\a_n^\nu]=[\ta_m^\mu,\ta_n^\nu]=m\delta_{m+n}\eta^{\mu\nu}, ~~
\{b_{r}^{\mu},b_{s}^{\nu} \}=\{\tilde{b}_{r}^{\mu},\tilde{b}_{s}^{\nu} \}=\delta_{r+s}\eta^{\mu\nu}. 
\ee
Anti-hermiticity of the $\psi_-$ implies $\tilde{b}^\dagger_r=-\tilde{b}_{-r}$. This in turn implies $\{\tilde{b}_r,\tilde{b}^\dagger_r\}=-1$. Now, by using the components of energy momentum tensor and the supercurrent, 
the constraints can be written down in terms of the mode expansions. Let us write here the antiholomorphic ones explicitly, which will be important for us later;
\bea{}\label{virgen}
\bar{\L}_n&=&\frac{1}{2}\sum_m \tilde{\a}_{-m}\cdot\tilde{\a}_{m+n}+\frac{1}{4}\sum_r (2r+n)\tilde{b}_{-r}\cdot \tilde{b}_{r+n}, 
\quad\bar{\Q}_r=\sum_m \tilde{\a}_{-m}\cdot\tilde{b}_{m+r}.
\eea
The anti-hermiticity of $\tilde b$ implies $\bar{\Q}_r$ is anti-hermitian. The constraints again generate the Super-Virasoro algebra \refb{algv}, and we see that that the anti-hermiticity does not affect the construction at the level of the algebra.

\paragraph{Tensionless limit on Worldsheet: }
\label{4.2}
The tensionless limit on the string worldsheet  \cite{Bagchi:2013bga} can be understood as a singular limit where the spatial co-ordinate $\s \to \infty$ keeping the worldsheet time $\tau$ fixed. Equivalently, in the bosonic case, this is: 
\be{url} \s\rightarrow\s\ \mbox{and} \ \t\rightarrow\eps\t,\ \ \eps\rightarrow 0.\ee
This amounts to taking the worldsheet speed of light effectively to zero. The limit is thus an ultra-relativistic or a \textit{Carrollian limit} on the worldsheet, where the bosonic generators of the Virasoro algebra change in the following way:
\be{iscal}
L_n=\L_n-\bL_{-n}, \quad M_n=\eps (\L_n+\bL_{-n}).
\ee
The symmetry algebra obeyed by these contracted generators is the 3d BMS (or equivalently the 2d GCA) \cite{Bagchi:2009pe}. Since we want the tensionless limit to be consistent, the string length $\a'$ should also scale as $\a'\rightarrow c'\eps$, where $c'$ is the finite parameter defined earlier. 

In the supersymmetric case, there are two possible scaling of the fermionic degrees of freedom. For the homogeneous contraction, all generic fermionic objects $\zeta$ scale as $ \zeta\rightarrow\sqrt{\eps}\zeta$. In the inhomogeneous case, two components of worldsheet spinors contract in different ways:
\be{} 
\psi_0\rightarrow\frac{1}{\eps}\psi_0,~~\psi_1\rightarrow\psi_1 
\ee
where the new spinors are linear combinations of tensile spinors: $\psi_{0,1}=\frac{1}{\sqrt{2}}(\psi_+ \pm i\psi_-)$. For consistency, the SUSY parameter $\e$ and the superspace coordinates $\theta,\bar{\theta}$ scale similarly:
\be{}
(\e^0, \e^1)=\frac{1}{\sqrt{2}}(\e^+ \mp i\e^-),\ \  (\a, \chi)=\frac{1}{\sqrt{2}}(\theta \mp i\bar{\theta});\quad (\e^0, \e^1) \rightarrow(\eps\e^0, \e^1); \, (\a, \chi) \rightarrow(\eps\a, \chi).
\ee
Starting from \refb{taction} we can take the UR limit at every step till \refb{virgen}. The bosonic and fermionic modes transform under this limit as 
\bea{}
X&=&
x+\sqrt{2c'}\sqrt{\eps}(\a_0+\ta_0)\tau+i\sqrt{2c'}\sum_{n\neq 0}\frac{1}{n}\Big[\frac{1}{\sqrt{\eps}}(\a_n-\ta_{-n})-i n\tau\sqrt{\eps}(\a_n+\ta_{-n})\Big]e^{-in\sigma},
\nonumber \\
\psi_0&=&\sqrt{c'}\sum_r \sqrt{\eps}(b_r+i\tilde{b}_{-r}) e^{-ir\sigma};~~
\psi_1=\sqrt{c'}\sum_r \Big[\frac{b_r-i\tilde{b}_{-r}}{\sqrt{\eps}}-ir\tau\sqrt{\eps}(b_r+i\tilde{b}_{-r})\Big]e^{-ir\sigma}. \nonumber
\eea
A comparison between the tensionless (\ref{NSsoln}) and the tensile oscillator modes gives us  
\bea{}
A_{n}={\eps}^{-1/2}(\a_n-\tilde{\a}_{-n}), B_{n}={\eps}^{1/2}(\a_n+\tilde{\a}_{-n}); \,
\gamma_{r}={\eps}^{-1/2}(b_r-i\tilde{b}_{-r}), \beta_{r}={\eps}^{1/2}(b_r+i\tilde{b}_{-r}).
\eea
The tensionless generators \refb{inhomgen} are related to that of the parent theory by 
\bea{ingenscal}
L_n=\L_n-\bL_{-n}, \, M_n = \eps (\L_n+\bL_{-n}), \quad
G_r=\Q_r-i\bQ_{-r}, \, H_r = \eps (\Q_r+i\bQ_{-r}).
\eea
Needless to say, it can be easily checked that these satisfy SBMS$_I$ \refb{agcai} and all of them are hermitian. Thus at a classical level, we have successfully found a parent theory from which we can consistently take the (Inhomogeneous) tensionless limit and find a theory with well defined oscillators and hermitian generators. 

\paragraph{Bogolioubov transformations and vacua:}
In \refb{sbasis}, we made a change of basis so that the commutation relations obey the usual harmonic oscillator algebra. In terms of the changed oscillators the above relation between tensile/tensionless modes now has a rather suggestive form:
\begin{subequations}\label{bogoll}
\bea{}
C_n&=&\a_n\cosh\varphi+\ta_{-n}\sinh\varphi,\quad \w_r=b_r\cosh\varphi+i\tb_{-r}\sinh\varphi; \\
\C_n&=&\a_{-n}\sinh\varphi+\ta_n\cosh\varphi,\quad \tw_r=-ib_{-r}\sinh\varphi+\tb_r\cosh\varphi,
\eea
\end{subequations}
where $\tanh\varphi=\frac{\eps-1}{\eps+1}$. These are \textit{Bogoliubov transformations on the worldsheet} parameterised by $\varphi$ since the canonical commutation relations are preserved in the course of this transformation. This gives us an explicit evolution of the tensile theory to its tensionless offspring by dialling $\varphi$ or equivalently $\eps$. The generators of the Bogoliubov transformations \refb{bogoll} for the bosonic and the fermionic parts can be written in the following form:
\begin{subequations}\label{gennn}
\bea{}
G_B(\varphi)&=&i\sum_{n>0} \varphi \Big[a_{-n}\cdot \tilde{a}_{-n}-a_n\cdot \tilde{a}_n\Big],~~
G_F(\varphi)=\sum_{r>0} \varphi \Big[b_{-r}\cdot \tilde{b}_{-r}-b_r\cdot \tilde{b}_r\Big], 
\eea
\end{subequations}
where $a_n=\frac{1}{\sqrt{n}}\a_n$ are normalised oscillator modes. It can be checked that both of these generators are hermitian in our case. In terms of these generators, the transformations the two sets of oscillators are given by:
\be{}
C_n=e^{-iG_B}\a_ne^{iG_B},\, \C_n=e^{-iG_B}\ta_ne^{iG_B}, \quad \w_r=e^{-iG_F}b_re^{iG_F}, \, \tw_r=e^{-iG_F}\tb_re^{iG_F}. 
\ee

In \cite{Bagchi:2015nca}, the bosonic version of this was considered. It was shown that the transformation implies a mapping between the tensionless vacuum $|0\>_C$ and the tensile one $|0\>_\a$ where one can be written as a squeezed state over the other.
\bea{}
|0\>_C=e^{iG_B(\varphi)}|0\>_\a = {\sqrt{\cosh\varphi}} \prod_{n>0} \exp\Big[\tanh\varphi\ a_{-n}\cdot\tilde{a}_{-n}\Big]|0\>_\a.
\eea
So the new tensionless vacuum is a highly energised state in comparison to the tensile vacuum. Notice that there is a normalisation constant which sits in front of the expression. When the tensionless limit is taken, this normalisation constant blows up, indicating something like a phase transition. In \cite{Bagchi:2019cay}, this was further explored and it was shown that in the tensionless limit, starting with closed strings one ends up with a Neumann boundary state, indicating that open strings (or space-filling D-branes) arise at the end point of this evolution. It was also conjectured that there is a Bose-Einstein like condensation on the worldsheet forming this open string from the closed string excitations. 

In the supersymmetric case here, using the generators \refb{gennn} the vacuum of the tensile and tensionless theories can again be mapped to one another as follows:
\be{}
 |00\>_C=e^{iG_B(\varphi)}e^{iG_F(\varphi)}|00\>_\a \nonumber\\
=\prod_{n,r>0} \exp\Big[\tanh\varphi(a_{-n}\cdot\tilde{a}_{-n}+ib_{-r}\cdot\tb_{-r})\Big]|00\>_\a. 
\ee
It is interesting to note that unlike in the bosonic case, there is no infinite normalisation factor outside the exponential \cite{Bagchi:2015nca} as we take the tensionless limit. In fact we can explicitly put $\eps=0$ in the above expression by recalling the relation between $\tanh\varphi$ and $\eps$, and obtain
\be{fvacc}
|00\>_C=\prod_{n,r>0} \exp\Big[-(a^\dagger_{n}\cdot\tilde{a}^\dagger_{n}-ib^\dagger_{r}\cdot\tb^\dagger_{r})\Big]|00\>_\a. 
\ee
The above supersymmetric state, in analogy to the bosonic one, is a highly squeezed state with respect to the $|00\>_\a$ vacuum. The above expression is in the form of a fermionic boundary state and there should be a similar story of the formation of an open string from closed strings that emerges in this context as well. But the absence of the infinite normalisation constant indicates that there would also be significant departures from the bosonic picture. We leave a detailed discussion of this for future work. 

\section{Exotic Parent II: Flipped Superstrings} 

In our first construction of a parent theory, we played with the hermiticity of one of the modes of the fermion to generate our tensionless theory. We now put forward a different theory, a  string theory with an unconventional vacuum state. 

\paragraph{An automorphism of the Super-Virasoro:}
Let us consider the following transformation on the anti-holomorphic sector of the tensile modes
\be{a1}
\ta_n \rightarrow \ta'_n=\pm i\ta_{-n},\quad\tb_r \rightarrow \tb'_r=\pm \tb_{-r}.
\ee
Here, the unprimed modes represent the usual hermitian Super-Virasoro modes, while the primed ones represent the changed modes. There is a clear \textit{flip} between the creation and annihilation operators.
 A bit of algebra gives us the associated transformation on the constraints
\be{auto}
\bar{\mathcal{L}}_n\rightarrow\bar{\mathcal{L}}'_n=-\bar{\mathcal{L}}_{-n};\quad\bar{\mathcal{Q}}_r\rightarrow\bar{\mathcal{Q}}'_r=\pm i\bar{\mathcal{Q}}_{-r};
\ee
Intriguingly, if we start from the usual Super-Virasoro constraints and perform the above transformation, the algebra would remain preserved. In the centrally extended version, the algebra will again remain invariant under above transformations provided we have, 
$ \bar{c} \to \bar{c}' = - \bar{c}.$
Hence this set of transformations is an \textit{automorphism} of the Super-Virasoro algebra. 

\paragraph{Tensile Flipped Vacuum:}
Without any loss of generality we can consider the case $\ta'_n=i\ta_{-n}$ and $\tb'_r=-\tb_{-r}$. Then we can see that the vacuum of the theory $|00\>_f$, is defined such that
\begin{subequations}\label{flipped}
\be{}
\a_n|00\>_f=b_r|00\>_f=0 ;~~~~  \ta_{-n}|00\>_f=\tb_{-r}|00\>_f=0\ \forall\ n,r>0.
\ee
\end{subequations}
The last line is condition for a \textit{flipped vacuum}, where the nature of the right-handed oscillators (and their actions on the vacuum) have been flipped. This asymmetric vacuum is reminiscent of works in the context of Ambitwistor string theory \footnote{For related introduction to the Ambitwistor strings, the reader should consult the seminal work \cite{Mason:2013sva} and references therein.}. A similar left-right asymmetric vacuum was first put forward briefly in \cite{Hwang:1998gs}, and later revisited in a series of works including \cite{Siegel:2015axg, Huang:2016bdd, Casali:2016atr}. A closed bosonic tensile string theory built on this vacuum was recently studied in \cite{Lee:2017utr}. 

\medskip
\noindent \textbf{The tensionless counterpart:} We saw that for the maverick tensile strings, the tensile vacuum evolved to the tensionless one via worldsheet Bogoliubov transformations. However, for the flipped case, the situation is quite different. Using the relations between the tensile and tensionless modes, and following the UR limit on the worldsheet, the condition for the flipped vacuum \refb{flipped} in terms of the oscillators $C_n$, $\C_n$, $\w_r$ and $\tw_r$ becomes 
\be{nvacc} C_n|00\>_F=\C_{-n}|00\>_F=\w_r|00\>_F=\tw_{-r}|00\>_F=0\ \ \  \forall\ n,r>0 \ee
This means that the vacuum $|00\>_F$ remains the vacuum in the tensionless theory. Of course, there may be pieces of the initial state that drop off in the tensionless limit. One way of envisioning such a case is by looking near the $\eps \to 0$ limit. Here we would be able to write
\be{}
|00\>_f = |00\>_F + \eps |01\> + \eps^2 |02\> + \ldots
\ee
The limit clearly gives us 
\be{}
\lim_{\eps \to 0} |00\>_f = |00\>_F
\ee
We can quantise the limiting theory around the same vacuum as the parent one. To this end, we will discuss the representation theory underlying this construction. 

\paragraph{Tensionless Representations:}
\label{HWR}
Once we define a vacuum of the theory, we can build up states by acting oscillator modes on it. We are going to define the physical string states appropriately, depending on the choice of vacua. Let us first start with the usual tensile case. The states are labelled by their conformal weights:  
\be{}
\L_0|h,\bar{h}\>= h |h,\bar{h}\>, \quad \bL_0|h,\bar{h}\>=\bar{h} |h,\bar{h}\>
\ee
A primary state $|h,\bar{h}\>_p$ is defined by 
\be{primary}
\L_n|h,\bar{h}\>=\bar{\L}_n|h,\bar{h}\>=0, \quad \Q_r|h,\bar{h}\>=\bar{\Q}_r|h,\bar{h}\>=0\ \forall\ n, r>0, 
\ee
The highest weight representations are built on the primary states by acting raising operators $\L_{-n}, \bar{\L}_{-n}, \Q_{-r}, \bar{\Q}_{-r}$. This builds the Hilbert space of the usual tensile string.   

\medskip 

\noindent Now, let us focus on the SBMS$_I$. We will define states by labelling them with $L_0$ and $M_0${\footnote{We work in the NS sector and so the SUSY generators are half-integer moded and hence don't contribute to the Cartan sub-algebra.}}. 
\be{}
L_0|h_L,h_M\>= h_L |h_L,h_M\>, \quad M_0|h_L,h_M\>= h_M |h_L,h_M\>.
\ee
The primary states are defined a way analogous to the above: 
\be{hwbms}
L_n|h_L,h_M\>_p=M_{n}|h_L,h_M\>_p=0, \quad G_r|h_L,h_M\>_p=H_{r}|h_L,h_M\>_p=0\ \forall\ n, r>0
\ee
The {\em{highest-weight representations}} of the SBMS$_I$ are built out of modules constructed in turn from these primary states by raising operators. 

\medskip 

\noindent We also define another class of representations, familiar from the Poincare algebra, the so-called {\em{induced representations}}. The states are still labelled by $L_0$ and $M_0$. 
\be{}
M_0|M, s\>_i= M |M, s\>_i, \quad L_0 |M, s\>_i= s |M, s\>_i.
\ee
But now the action of the generators for non-zero modes on the states is different. These are given by 
\be{irbms}
M_{n}|M, s\>_i=0, \quad H_{r}|M, s\>_i=0\ \forall\ n, r \neq 0.
\ee
Generic states are constructed by acting on these states with $L_n$ and $G_r$ for all positive and negative values of the modes. For more details, the readers are pointed to \cite{Barnich:2014kra, Campoleoni:2015qrh} for a discussion of the bosonic algebra and the homogeneous supersymmetric algebra. 

\paragraph{Representations, Limit and Flip:} Let us rewrite \refb{ingenscal} in a form that will be helpful in understanding the action of the limit on the representation theory.
\bea{}
2 \L_n = \frac{1}{\eps} M_n + L_n, \, 2\bL_n =\frac{1}{\eps} M_{-n} - L_{-n}; \quad 2\Q_r = G_r+ \frac{1}{\eps} H_r, \, 2\bQ_r = i \left(G_{-r} - \frac{1}{\eps} H_{-r} \right)
\eea
It is clear from the above that in the UR limit, the primary state conditions of Super-Virasoro \refb{primary} {\em{do not}} map to primary conditions of SBMS$_I$ \refb{hwbms} since there is a mixing of positive and negative modes of the generators in the UR limit. The highest weight representations of the Super-Virasoro, in the limit, then do not yield highest-weight representations of the SBMS$_I$, but instead lead to the induced representation \refb{irbms}:
\bea{}
&& \L_n |\varphi\> = 0 \, \Rightarrow M_n |\phi\> = 0, \quad \bL_n |\varphi\> = 0 \, \Rightarrow M_{-n} |\phi\> = 0, \quad \forall n>0; \nonumber\\
&& \Q_r |\varphi\> = 0 \, \Rightarrow H_r |\phi\> = 0, \quad \bQ_r |\varphi\> = 0 \, \Rightarrow H_{-r} |\phi\> = 0, \quad \forall r >0. 
\eea
Notice that the labels on the state, however, do map to each other. The labels are related by 
\be{}
|\varphi\> = |h, \bar{h}\>, \, \, |\phi\> = |M, s\>, \quad \mbox{where} \quad M= \eps(h + \bar{h}), \, s = h-\bar{h}. 
\ee

On the other hand, there exists another limit from the Super-Virasoro algebra to the SBMS$_I$ that can be thought of as a non-relativistic (NR) limit where the speed of light goes to infinity instead of zero. The generators in this limit scale in the following fashion
\be{NRscal}
L_n=\L_n+\bL_{n}, \quad M_n = \eps (\L_n-\bL_{n}), \quad G_r=Q_r+\bQ_{r}, \quad H_r = \eps (Q_r-\bQ_{r}).
\ee
Notice now that there is no mixing between positive and negative modes in this limit. Hence, following steps outlined above, we can see that the NR limit would yield highest-weight representations of SBMS$_I$ when you start with the Super-Virasoro highest-weights. The weights are now related by 
\be{}
 h_L = h+\bar{h}, \, h_M= \eps(h - \bar{h}).
\ee
Notice that the two contractions \refb{ingenscal} and \refb{NRscal} are related by the automorphism of the Virasoro algebra \refb{auto} discussed at the beginning of the section{\footnote{This seems to be at the heart of recent discussions of the isomorphism between NR and UR conformal algebras discussed in the context of 3d Minkowskian holography \cite{Bagchi:2010eg}. Some pertinent comments regarding the bosonic version of this automorphism can be found in  \cite{Bagchi:2019unf}.}}.

In the case of the flipped vacuum, starting with the conditions \refb{flipped}, we can once again construct primary states, with a sign on the modes of the anti-holomorphic oscillators. We will need to modify \refb{primary} by the following expression
\begin{subequations}{}
\bea{}
\L_n|h,\bar{h}\>=\bar{\L}_{-n}|h,\bar{h}\>=0,&&
\Q_r|h,\bar{h}\>=\bar{\Q}_{-r}|h,\bar{h}\>=0\ \forall\ n, r>0, \\
(\L_0-h)|h,\bar{h}\>&=&(\bL_0+\bar{h})|h,\bar{h}\>=0.
\eea
\end{subequations}
This effectively means that we are imposing different representations of the super-Virasoro on left and right handed sides. We have one highest weight representation tensored with a lowest weight representation. 

Applying the contraction as in \refb{ingenscal}, we see that for the UR limit on the ``flipped'' vacuum effectively acts as a NR limit. We adapt the mantra:
\be{}
\mbox{Flipped} \oplus \mbox{UR limit} = \mbox{NR limit}
\ee
Under the UR limit, the primaries of Super-Virasoro in this flipped sector map to the primaries of the SBMS$_I$. These states are now 
\begin{subequations}{}
\bea{}
L_n|h_L,h_M\>=M_{n}|h_L,h_M\>=0, &&
G_r|h_L,h_M\>=H_{r}|h_L,h_M\>=0\ \forall\ n, r>0, \\
(L_0-h_L)|h_L,h_M\>&=&(M_0-h_M)|h_L,h_M\>=0.
\eea
\end{subequations}
In the above, $h_L=h-\bar{h}$ and $h_M=\eps(h+\bar{h})$. The vacuum of the tensionless limit of this theory would satisfy the above relations. Therefore, just by tweaking one sector, not only will we have well-defined hermitian constraints, but also a highest weight representation of SBMS$_I$. 

\paragraph{Connections with Ambitwistors}
 
It is rather intriguing to note that tracing back the origins of inhomogeneous tensionless strings naturally gives rise to such flipped string vacua as a consequence. The recent connection between the ambitwistor string and the tensionless limit proposed in \cite{Casali:2016atr} asserted that the ambitwistor strings arose from an alternative gauge choice on the tensionless worldsheet theory. The fact that a flipped vacuum also exists for the ambitwistor string points to two interesting consequences. Firstly, we note that the symmetry of the ambitwistor superstring in its present avatar is governed by the SBMS$_H$. Our current construction hints at a new \textit{inhomogeneous counterpart of the ambitwistor superstring}. Secondly, we have identified a parent theory and this should be very useful in the endeavour of constructing the origin of the ambitwistor superstring. 

Let us add something more about why the inhomogeneous algebra would be a good candidate for the ambitwistor superstring. The ambitwistor string is a chiral object and the bosonic version is governed by a single Virasoro algebra. This is so even though the symmetry that arises on the worldsheet is BMS$_3$. By considering the analysis of null vectors in the highest weight representations performed in \cite{Bagchi:2009pe}, which reduces the BMS$_3$ to its Virasoro subalgebra, \cite{Casali:2017zkz} proposed how to reconcile the worldsheet symmetries of the ambitwistor string to its chiral nature. The SBMS$_I$ contains a single Super-virasoro sub-algbera, unlike the SBMS$_H$, and one can perform a truncation of the SBMS$_I$ down to this subalgebra in the case where the central charge $c_M$ vanishes \cite{Bagchi:2018ryy, Bagchi:2017cte}. It is thus very likely that there would exist a supersymmetric version of the Ambitwistor string which has the SBMS$_I$ as its underlying algebra and the chiral nature of ambitwistor superstring would manifest itself in a way very similar to the bosonic case.

\section{Conclusions}

In this paper, we revisited the construction of the inhomogeneous tensionless superstring introduced in \cite{Bagchi:2017cte}. We demanded reality of spinors and this led us to manifestly hermitian generators for the underlying residual symmetry algebra, the SBMS$_I$. We then attempted to answer what properties of a tensile theory this version of the tensionless superstring arose from and found two distinct ways of implementing this. The first was the Maverick superstring where fermions in the parent theory has certain intriguing properties. Armed with this knowledge, as a natural extension of the bosonic analysis in \cite{Bagchi:2015nca}, we constructed the Bogoliubov transformations that connect the tensile and tensionless oscillators and thereby came up with the mapping between the two superstring  vacua. The other parent tensile superstring, which we called the Flipped superstring was one with an unconventional vacuum. Our construction was  based on a discrete automorphism of the parent Super-virasoro algebra. This observation led us to this different vacuum, which has been considered recently in \cite{Casali:2016atr} in the context of Ambitwistor strings. We made several comments on the underlying representation theory and possible emergence of a new ambitwistor superstring. 

\subsection*{Acknowledgements}

We thank Punit Sharma for early collaboration on aspects of this work. We also thank Rudranil Basu, David Skinner and Wei Song for useful discussions. This research is supported by the following grants: DST-Inspire faculty fellowship (Arjun Bagchi [AB]), Mathematical Research Impact Centric Support by SERB (AB), the Max Planck visiting fellowship (AB, PP), the Chinese Academy of Sciences (CAS) Hundred-Talent Program (Aritra Banerjee [ArB]), Key Research Program of Frontier Sciences, CAS (ArB), Project 11647601 supported by NSFC (ArB), ISIRD grant 9-252/2016/IITRPR/708 (SC), Grant No. 09/936(0098)/2013-EMR-I, SERB (PP). 
%
%
%

\bibliographystyle{JHEP}
\bibliography{ref.bib}

%
%
\end{document}